\documentclass[letterpaper]{article}

\usepackage{natbib}  

\title{\vspace{-1cm}Co-evolution and morphogenetic systems}
\author{Juste Raimbault$^{1,2}$\\
\mbox{}\\
$^1$UPS CNRS 3611 ISC-PIF, Paris, France \\
$^2$UMR CNRS 8504 G{\'e}ographie-cit{\'e}s, Paris, France \\
juste.raimbault@polytechnique.edu} 

\date{}

\begin{document}

\maketitle

\begin{abstract}
The emerging field of morphogenetic engineering proposes to design complex heterogeneous system focused on the paradigm of emergence. Necessarily at the interface of disciplines, its concepts can be defined through multiple viewpoints. This contribution aims at linking a co-evolutionary perspective on such systems with morphogenesis, and therein at bringing a novel conceptual approach to the bottom-up design of complex systems which allows to fully consider co-evolutive processes. We first situate systems of interest at the interface between biological and social systems, and introduce a multidisciplinary perspective on co-evolution. Building on Holland's signals and boundaries theory of complex adaptive systems, we finally suggest that morphogenetic systems are equivalent to combinations of co-evolutionary niches. This introduces an entry to morphogenetic engineering focused on co-evolution between components of a system. Applications can be found in a broad range of subjects, which we illustrate with the example of planning in territorial systems, suggesting an extended scope for the relevance of morphogenetic engineering concepts.
\end{abstract}

\section{Introduction}

The standard opposition between bottom-up self-organized systems and top-down architected has been challenged by the field of morphogenetic engineering (ME), which focuses on the design of systems from the bottom-up: tuning the behavior of agents and processes at the micro level allows in this paradigm to obtain a desired functionality at the macro level, without any centralization \citep{doursat2013review}. \cite{doursat2011myriads} identifies two streams of research embracing this approach, namely complex systems controlled by Information and Communication Technologies (ICT) and ICT inspired by complex systems, typically in the spirit of bio-inspired designs. Swarm chemistry \citep{sayama2009swarm} is a good illustration how simple rules produce complex emergent patterns in swarm of particles, and how these can be interactively evolved \citep{sayama2009enhancing}. The type of systems to which ME applies stays however limited, as it for example does not grasp social or territorial systems. The theoretical and practical implications of an extension in scope of ME would be significant, as it would for example converge with bottom-up approach to planning for territorial systems \citep{batty2007cities}. 

We propose here such a extension in scope, in an indirect way through the conceptual exercise of a new entry on morphogenetic systems from the point of view of co-evolution. The aim of this paper is thus twofold: (i) develop the concept of co-evolution and show how it connects to morphogenesis and ME; (ii) show how this link between co-evolution and morphogenesis opens new potentialities for the application of ME concepts. The rest of the paper is organized as follows: we first develop the parallel between biological and social systems, what will suggest the exploration of the concept of co-evolution, on which a multidiscplinary view is then constructed. We recall then the concept of morphogenesis and introduce its connexion with co-evolution. This allows us to conclude with an illustration of a possible scope extension of ME to territorial systems.

\section{A parallel between biological and social systems}

\subsection{Epistemology of life}

The parallel between social and biological systems is not rare, altough it sometimes relies on analogies as for example in West's \emph{Scaling} theory which applies similar growth equations starting from scaling laws, with however inverse conclusions concerning the relation between size and pace of life~\citep{bettencourt2007growth}. Scaling relations do not hold when we try to apply them to a single ant, and they must be applied to the whole ant colony which is then the organism studied. When adding the property of cognition, we confirm that it is the relevant level, since the colony shows advanced cognitive properties, such as the resolution of spatial optimization problems, or the quick answer to an external perturbation. Human social organizations, cities, could be seen as organisms ? \cite{banos2013pour} extends the metaphor of the \emph{urban anthill} but recalls that the parallel stops quickly. We can however show that some concepts from the epistemology of biology can be useful to understand social systems.

\cite{monod1970hasard} develops epistemological principles for the study of life. Living organisms answer to three essential properties that differentiate them from other systems: (i) the teleonomy , i.e. the property that these are ``objects with a project'', project that is reflected in their structure and the structure of artifacts they produce; (ii) the importance of morphogenetic processes in their constitution; (iii) the property of the invariant reproduction of information defining their structure. Monod furthermore introduces research directions towards a theory of cultural evolution. Teleonomy is crucial in social structures, since any organization aims at satisfying a set of objectives, even if in general it will not succeed and the objectives will co-evolve with the organization. This notion of multi-objective optimization is typical of complex socio-technical systems.


\subsection{Morphogenesis and co-evolution}

We postulate that the concept of morphogenesis is an essential tool to understand social systems, with a definition very similar to the one used in biology. We sum it up as the existence of relatively autonomous processes guiding the growth of the system and implying causal circular relations between form and function, that witness an emergent architecture in the sense of \cite{doursat2012morphogenetic}. For social systems, isolating the system is more difficult and the notion of boundary will be less struct than for a biological system. 

Finally, the reproduction of information is at the core of cultural evolution, through the transmission of culture and \emph{memetics}, the difference being that the ratio of scales between the frequency of transmission and mutation and cross-over processes or other non-memetic processes of cultural production is relatively low, whereas is many orders of magnitude in biology. An example shows that the parallel is not always absurd : \cite{2017arXiv170305917G} proposes an auto-catalytic network model for cognition, that would explain the apparition of cultural evolution through processes that are analogous to the ones that occurred at the apparition of life. 
But even if processes are at the origin analogous, the nature of evolution is then quite different, as show \cite{vanderLeeuw2009}, Darwinian criteria for evolution being not sufficient to explain the evolution of our organized societies.

\subsection{Levels of emergence}

One point that also must retain our attention is the greater difficulty to define levels of emergence for social systems: \cite{roth2009reconstruction} underlines the risk to fall into ontological dead-ends if levels were badly defined. He argues that more generally we must go past the single dichotomy micro-macro that is used as a caricature of the concepts of weak emergence, and that ontologies must often be multi-level and imply multiple intermediate levels. This echoes to the question of the existence of strong emergence in social structures, that in sociological terms corresponds to the idea of the existence of ``collective beings''~\citep{angeletti2015etres}. \cite{morin1980methode} indeed distinguishes living systems of the second type (multi-cellular) and of the third type (social structures), but precises that the \emph{subjects} of the latest are necessarily unachieved.

\medskip

This thematic entry on social and biological systems suggests therefore a crucial conceptual role of morphogenesis and co-evolution in understanding these systems, and therefore a possible link between morphogenetic systems and co-evolution. We will now develop this concept in more details.

\section{A multidisciplinary approach to co-evolution}

Co-evolution is indeed used in several disciplines, but does not have a unified definition. We propose to take advantage of this disparity of views and construct a multidisciplinary approach. We sequentially review several disciplines, starting from biology with a progressive shift to social systems.

\subsection{Biology}

The concept of co-evolution in biology is an extension of the well-known concept of \emph{evolution}, that can be tracked back to Darwin. \cite{durham1991coevolution} recalls the components and systemic structures that are necessary for the presence of evolution, that are:

\begin{enumerate}
\item Processes of \emph{transmission}, implying transmission units and transmission mechanisms.
\item Processes of \emph{transformation}, that necessitates sources of variation.
\item Isolation of sub-systems such that the effects of previous processes are observable in differentiations.
\end{enumerate}

Therefore, a population submitted to constraints (often conceptually synthesized as a \emph{fitness}) that condition the transmission of the genetic heritage of individuals (transmission), and to random genetic mutations (transformation), will indeed be in evolution in the spatial territories it populates (isolation), and by extension the species to which it can be associated. 

Co-evolution is then defined as an evolutionary change in a characteristic of individuals of a population, in response to a change in a second population, which in turn responds by evolution to the change in the first, as synthesized by~\cite{janzen1980coevolution}. He furthermore highlights the subtlety of the concept and warns against its unjustified uses: the presence of a congruence between two characteristics that seem adapted one to the other does not necessarily imply a co-evolution, since one species could have adapted alone to one characteristic already present in the other.

This rough presentation partly hides the real complexity of ecosystems: populations are embedded in trophic networks and environments, and co-evolutionary interactions would imply communities of populations from diverse species, as presented by \cite{strauss2005toward} under the appellation of diffuse co-evolution. Similarly, spatio-temporal dynamics are crucial in the realization of these processes: \cite{dybdahl1996geography} study for example the influence of the spatial distribution on patterns of co-evolution for a snail and its parasite, and show that a higher speed of genetic diffusion in space for the parasite drive the co-evolutionary dynamics.

The essential concepts to retain from the biological point of view are thus: (i) existence of evolution processes, in particular transmission and transformation; (ii) in circular schemas between populations in the case of co-evolution; and (iii) in a complex territorial frame (spatio-temporal and environmental in the sense of the rest of the ecosystem).

\subsection{Cultural evolution}

This development on co-evolution was brought by the parallel between biological and social systems. The evolution of culture is theorized within a proper field, and witnesses many co-evolutive dynamics. \cite{Mesoudi25072017} recalls the state of knowledge on the subject and future issues, such as the relation with the cumulative nature of culture, the influence of demography in evolution processes, or the construction of phylogenetic methods allowing to reconstruct branches of past evolutionary trees.

To give an example, \cite{carrignon2015modelling} introduces a conceptual frame for the co-evolution of culture and commerce in the case of ancient societies for which there are archeological data, and proposes its implementation with a multi-agent model which dynamics are partly validated by the study of stylized facts produced by the model. The co-evolution is here indeed taken in the sense of a mutual adaptation of socio-spatial structures, at comparable time scales, in this more general frame of cultural evolution.

Cultural evolution would even be indissociable from genetic evolution, since \cite{durham1991coevolution} postulates and illustrates a strong link between the two, that would themselves be in co-evolution. \cite{bull2000meme} explore a stylized model including two types of replicant populations (genes and memes) and shows the existence of phase transitions for the results of the genetic evolution process when the interaction with the cultural replicant is strong.

\subsection{Sociology}

The concept was used in sociology and related disciplines such as organisation studies, following the parallel done before the same way as cultural evolution. In the field of the study of organisations, \cite{volberda2003co} develop a conceptual frame of inter-organisational co-evolution in relation with internal management processes, but deplore the absence of empirical studies aiming at quantifying this co-evolution. In the context of production systems management, \cite{tolio2010species} conceptualize an intelligent production chain where product, process and the production system must be in co-evolution.

\subsection{Economic geography}

In economic geography, the concept of co-evolution has also largely been used. The idea of evolutionary entities in economy comes in opposition to the neo-classical current which remains a majority, but finds a more and more relevant echo~\citep{nelson2009evolutionary}. \cite{schamp201020} proceeds to an epistemological analysis of the use of co-evolution, and opposes the view of a neo-schumpeterian approach to economy which considers the emergence of populations that evolve from micro-economic rules (what would correspond to a direct and relatively isolationist reading of biological evolution) to a systemic approach that would consider the economy as an evolutive system in a global perspective (what would correspond to diffuse co-evolution that we previously developed), to propose a precise characterization that would correspond to the first case, assuming co-evolving \emph{institutions}. The most important for our purpose is that he underlines the crucial aspect of the choice of populations and of considered entities, of the geographical area, and highlights the importance of the existence of causal circular relations.

Diverse examples of application can be given. \cite{colletis2010co} introduces a framework for the co-evolution of territories and technology (questioning for example the role of proximity on innovations), that reveals again the importance of the institutional aspect. The framework proposed by \cite{ter2011co} couples the evolutionary approach to companies, the literature on industries and innovation in clusters, and the approach through complex networks of connexions between the latest in the territorial system. In environmental economics, \cite{kallis2007coevolution} show that ``broad'' approaches (that can consider most of co-dynamics as co-evolutive) are opposed to stricter approaches (in the spirit of the definition given by \cite{schamp201020}).

\subsection{Geography}

In geography, the stream of research which is the closest empirically and theoretically to notions of co-evolution corresponds to the evolutive urban theory. The concept of co-evolution was present since the foundations of the theory \citep{pumain1997pour}: the system of cities as a complex adaptive system is composed of subsystems that are interdependent in a complex way, often with circular causalities. The corrsponding models include this vision in an implicit way, but co-evolution is not explicitly highlighted nor precisely defined, in terms that would be quantifiable or structurally identifiable. \cite{paulus2004coevolution} gives empirical proofs of mechanisms of co-evolution through the study of the evolution of economic profiles of French cities. The interpretation used by~\cite{schmitt2014modelisation} is based on an entry by the evolutive urban theory, and fundamentally consists in a reading of systems of cities as highly interdependent entities.

\subsection{Physics}

Finally, we can mention that the term of co-evolution is also used in physics. Its use for physical systems may induce some debates, depending if we suppose or not that the transmission assumes a transmission of \emph{information}\footnote{Information is defined within the shanonian theory as an occurence probability for a chain of characters. \cite{morin1976methode} shows that the concept of information must indeed be thought conjointly to a given context of the generation of a self-organizing negentropic system, i.e. realizing local decreases in entropy in particular thanks to this information. This type of system is necessarily alive. We consider here this complex approach to information.}. In the case of a purely physical ontological transmission (\emph{physical beings}), then a large part of physical systems are evolutive. \cite{hopkins2008cosmological} develop a cosmological frame for the co-evolution of cosmic heterogenous objects which presence and dynamics are difficultly explained by more classical theories (some types of galaxies, quasars, supermassive black holes). \cite{antonioni2017coevolution} study the co-evolution between synchronization and cooperation properties within a Kuramoto oscillators network.

\subsection{A synthetic definition of co-evolution}

Most of these approaches fit in the theory of complex adaptive systems developed by~\cite{holland2012signals}: it takes any system as an imbrication of systems of boundaries, that filter signals or objects. Within a given limit, the corresponding subsystem is relatively autonomous from the outside, and is called an \emph{ecological niche}, in a direct correspondence with highly connected communities within trophic or ecological networks. This way, interdependent entities within a niche are said to be co-evolving. We will come back on that approach below, when establishing the link with morphogenesis.

We retain from this multidisciplinary view of co-evolution the following important points:

\begin{enumerate}
	\item The presence of \emph{evolution processes} is primary, and their definition is almost always based on the existence of transmission and transformation processes.
	\item Co-evolution assumes entities or systems, belonging to distinct classes, which evolutive dynamics are coupled in a circular causal way. Approaches can differ depending on the assumptions of populations of these entities, singular objects, or components of a global system then in mutual interdependency without a direct circularity.
	\item The delineation of systems and subsystems, both in the ontological space (definition of studied objects), but also in space and time, and their distribution in these spaces, is fundamental for the existence of co-evolutionary dynamics, and it seems in a large number of cases, of their empirical characterization.
\end{enumerate}

We propose therefore a synthetic entry for the specific case of transportation networks and territories, echoing these points. It verifies the three following specifications. First of all, evolutive processes correspond to transformations of components of the system at the different scales: for the example of territorial systems transformation of cities on the long time, of their networks, transmission between cities of socio-economic characteristics carries by microscopic agents but also cultural transmission, reproduction and transformation of agents themselves (firms, households, operators)\footnote{This list is based on assumptions of the evolutive urban theory developed by \cite{pumain2008socio}. It can not be exhaustive, since what would be the ``ADN of a city'' remains an open question.}.

These evolutive processes may imply a co-evolution. Within a territorial system, can simultaneously co-evolve: (i) given entities (a given infrastructure and given characteristics of a given territory for example, i.e. individuals), when their mutual influence will be circularly causal (at the corresponding scale); (ii) populations of entities, what will be translated for example as such type of infrastructure and given territorial components co-evolve at a statistical level in a given geographical region; (iii) all the components of a system when there exists strong global interdependencies at a large scale. Our approach is thus fundamentally \emph{multi-scale} and articulates different significations at different scales.

Finally, the constraint of an isolation implies, in relation with the previous point, that co-evolution and the articulation of significations will have a meaning if there exists spatio-temporal isolations of subsystems in which different co-evolutions operate.

\section{Morphogenesis}

Morphogenesis is at the very heart of concepts in ME. We have furthermore seen that it naturally arises when comparing biological and social systems. We will precise now its meaning and unveil its link with co-evolution.

\subsection{Definition of morphogenesis}

The morphogenesis of a system consists in evolution rules that produce the emergence of its successives states and its form. It corresponds to self-organization, with the additional property that an \emph{emergent architecture} exists \citep{doursat2012morphogenetic}, in the sense of causal circular relations between the form and the function.

Morphogenesis can be tracked back to first contributions in biology such as \citep{thompson1942growth}, but the first interdisciplinary perspective can be with few doubts attributed to \cite{Turing1952} which introduced a reaction-diffusion model to explain the growth of the embryo. Progresses towards the understanding of embryo morphogenesis (in particular the isolation of particular processes producing the differentiation of cells from an unique cell) have been made only recently with the use of complexity approaches in integrative biology~\citep{delile2016chapitre}.

To show how morphogenesis applies to social systems, we can consider the example of urban systems. The idea of urban morphogenesis, i.e. of self-consistent mechanisms that would produce the urban form, is used in the field of architecture and urban design (as for example the generative grammar of ``Pattern Language'' of \cite{alexander1977pattern}), in relation with theories of urban form~\citep{moudon1997urban}. This idea can be pushed to small spatial extents such as the one of the building~\citep{whitehand1999urban} but it is in our case more relevant to consider it at a mesoscopic scale, in terms of land-use changes within an intermediate scale of territorial systems, with similar ontologies as the urban morphogenesis modeling literature. For example \cite{bonin2012modele} describes a model of urban morphogenesis with qualitative differentiation, whereas \cite{makse1998modeling} give a model of urban growth based on a mono-centric population distribution perturbed with correlated noises. \cite{raimbault2014hybrid} show with a model of co-evolution between transportation networks and housing how the interplay between the urban form and functions plays a determining role in the emergent behavior of the urban system.

The concept of morphogenesis is important in our approach in link with modularity and scale. Modularity of a complex system consists in its decomposition into relatively independent sub-modules, and the modular decomposition of a system can be seen as a way to disentangle non-intrinsic correlations~\citep{2015arXiv150904386K} (what can be thought in analogy to a block diagonalisation of a first order dynamical system). In the context of large-scale cyber-physical systems design and control, similar issues naturally raise and specific techniques are needed to scale up simple control methods~\citep{2017arXiv170105880W}. The isolation of a subsystem yields a corresponding characteristic scale. Isolating possible morphogenesis processes implies a controlled extraction (controlled boundary conditions e.g.) of the considered system, corresponding to a modularity level and thus a scale. When local processes are not enough to explain the evolution of a system (with reasonable variations of initial conditions), a change in scale is necessary, caused by an underlying phase transition in modularity.

Finally, it is important to remark that a subsystem for which morphogenesis makes sense, which boundaries are well defined and which processes allow it to maintain itself as a network of processes, is close to an \emph{auto-poietic system} in the extended sense of~\cite{bourgine2004autopoiesis}\footnote{Which are however not cognitive, making these morphogenetic systems not alive in the sense of auto-poietic and cognitive. Given the difficulty to define the delineation of cities for example, we will leave open the issue of the existence of auto-poietic territorial systems, and will consider in the following a less restrictive point of view on boundaries.}. These systems regulate then their boundary conditions, what underlines the importance of boundaries that we will finally develop.

\subsection{Linking morphogenesis and co-evolution}

We can now introduce the link between morphogenesis and co-evolution. It is build on Holland's approach to complex adaptive systems (CAS) with his theory of CAS as agents which fundamental property is to process signals through to their boundaries~\citep{holland2012signals}. In this theory, complex adaptive systems form aggregates at diverse hierarchical levels, which correspond to different level of self-organization, and boundaries are vertically and horizontally entangled in a complex way. That approach introduces the notion of \emph{niche} as a relatively independent subsystem in which resources circulate (the same way as communities in a network): numerous illustrations such as economical niches or ecological niches can be given. Agents within a niche are then said to be \emph{co-evolving}.

This notion of co-evolution coincides with the definition we gave, in particular at the level of a population of entities. The decomposition of a system in niches relates then to the modular decomposition that came in the morphogenesis approach. We postulate therefore the following fundamental conceptual link: \emph{the presence of morphogenetic processes in a system is equivalent to a decomposition in niches in which its components are co-evolving}. In other words, the properties of a morphogenetic system are essentially determined by the co-evolutive processes between its elements.

This proposition does not fundamentally changes the way to consider morphogenetic engineering, but gives a complementary entry, with strong basis in biology and other disciplines. We believe that this epistemological loop demonstrates the relevance of ME to study and design such kind of systems, since it exhibits a correspondence between the core concept of morphogenesis and an approach that is essentially multidisciplinary. This legitimates an extent of the scope of ME, as we will finally illustrate for the case of planning of territorial systems.

\section{An example of application: planning of territorial systems as morphogenetic engineering}

Territorial and urban planning correspond to a well established tradition of top-down approach to the design of systems. It has been shown by \cite{hall1982great} that this approach is far from being optimal both in terms of correspondence to the expected evolution of the corresponding territorial system, but also in terms of performance indicators. On the other hand, \cite{jane1961death} early suggested that bottom-up processes and neighborhood effects were crucial for the well-being of a city, close to seeing it as a morphogenetic living organism. Planning practices remain however highly architectured and hierarchical.

Moreover, the evolutive urban theory \citep{pumain1997pour} suggests that territorial systems are mostly composed of co-evolutive entities. In particular, the interactions and strong interdependencies between cities would be significant drivers of their growth processes~\citep{doi:10.1177/0042098010377366}, through the hierarchical diffusion of innovation which cycles are furthermore a potential explanation for urban scaling laws \citep{pumain2006evolutionary}. This type of system integrates co-evolution niches both as system components (the set of co-evolving agents) but also as the spatial support of the system (the regional territories that are relatively modular one to another).

The importance of co-evolutive processes in territorial systems and the conceptual link we established above suggest the relevance of applying morphogenesis to territorial systems: considering the appropriated co-evolutive niches, what would be equivalent to the appropriated morphogenetic subsystems, a problem of territorial planning can be mapped to a problem of ME. It would allow to answer questions such as how to obtain a desired territorial functionality from the bottom-up, by local actions on the behavior of agents or local interventions on the urban environment, what is in fact the exact logic of ME applied to the planning of territorial systems. We claim that this transfer could not be made directly, as the notion of morphogenesis of territorial systems is fuzzy and ontologies, scales and boundaries of the considered system would be difficult to consider without the conceptual entry of co-evolution.

There naturally remain unanswered questions on the practical implementation of such a bottom-up ME approach to territorial planning, since similar planning approaches applying complex systems paradigms such as fractal cities by \cite{frankhauser2018integrated} still keep a high level of top-down control. These questions should be the object of future work that can be investigated for example through modeling.

This illustration shows how this conceptual extension of ME by linking morphogenesis and co-evolution allows the application of ME to new types of systems, in particular territorial systems that are a given case of socio-technical systems. We believe broader applications should be found, since an action on the theoretical domain of knowledge (here a conceptual development) must impact the empirical domain (here the fields of application) because of their close relation~\citep{raimbault2017applied}.

\section{Conclusion}

In conclusion, a parallel between biological and social systems done in the spirit of extending the scope of ME has lead us to a multidisciplinary approach to co-evolution. We then linked it with morphogenesis, suggesting an entry to morphogenetic systems from the point of view of co-evolution. The application to the planning of territorial systems as morphogenetic design shows the potentialities of this conceptual extension. Our contribution is significant since it is to the best of our knowledge the first time that (i) an approach to co-evolution embracing a broad spectrum of disciplines is introduced; (ii) an explicit link between morphogenesis and co-evolution is proposed.


\footnotesize

\end{document}